\documentclass[prb,aps,twocolumn,dcolumn,superscriptaddress,showpacs,amsfonts,amsmath,amssymb,showkeys,floatfix]{revtex4-1}
\usepackage{bm}
\usepackage{pstricks}
\usepackage{setspace}
\usepackage{graphicx}% Include figure filesf
\begin{document}
\title{Nonlinear and  spin-glass susceptibilities of three site-diluted systems}
\date{\today}
\author{Julio F. Fern\'andez}
\affiliation{Instituto de Ciencia de Materiales de Arag\'on, CSIC-Universidad de Zaragoza, 50009-Zaragoza, Spain}
\affiliation{Instituto Carlos I de F\'{\i}sica Te\'orica y Computacional,  Universidad de Granada, 18071 Granada, Spain}
%\author{Juan J. Alonso}
%\affiliation{F\'{\i}sica Aplicada I, Universidad de M\'alaga,
%29071-M\'alaga, Spain}
%\altaffiliation{IVIC}
%\email[E-mail address: ] {}

\date{\today}
%\keywords

\begin{abstract}
The nonlinear magnetic  $\chi_{3}$ and spin-glass $\chi_{sg}$ susceptibilities in zero applied field are obtained, from tempered Monte Carlo simulations, for three different spin glasses (SGs) of Ising spins with quenched site disorder.
We find  that the relation $-T^3\chi_3=\chi_{sg}-2/3$  ($T$ is the temperature), which holds for Edwards-Anderson SGs, is approximately fulfilled in 
canonical-like SGs. For nearest neighbor antiferromagnetic interactions, on a $0.4$ fraction of all sites in fcc lattices, as well as for  spatially disordered Ising dipolar (DID) 
systems, $-T^3\chi_3$ and $\chi_{sg}$ appear to diverge  in the same manner at the critical temperature $T_{sg}$. However,  $-T^3\chi_3$ is  smaller than $ \chi_{sg}$  by over two orders of magnitude in the diluted fcc system. In DID systems,  $-T^3\chi_3/\chi_{sg}$ is very sensitive to the systems aspect ratio. Whereas near $T_{sg}$, $\chi_{sg}$ varies by approximately a factor of $2$ as system shape varies from  cubic to long-thin-needle shapes, $\chi_3$ sweeps over some four decades.
\end{abstract}

\maketitle

\section{Introduction}

The existence of an equilibrium phase transition into the spin glass (SG) phase has not yet been convincingly established for some spin glasses. The development of the \emph{parallel tempered} Monte Carlo (TMC) algorithm  \cite{TMC} has enabled one to observe, bypassing anomalously long relaxation processes, SG models in \emph{equilibrium} at low temperatures. Thus,  correlation lengths $\xi$ have been determined from the equilibrium behavior of $\langle \langle s_is_j \rangle_T ^2 \rangle_q$, where $s_i=\pm 1$ is for a spin at site $i$, 
and $\langle \cdots \rangle_T$ and $\langle \cdots \rangle_q$ stand for a thermal average and for an average over quenched randomness, respectively. There is evidence, from Monte Carlo simulations, that $\xi$ grows as linear system size $L$ in (i) the Edwards-Anderson (EA) model\cite{balle,katz} at some nonzero temperature $T_{sg}$ in three dimensions, in  (ii)  geometrically frustrated systems, such as strongly site-diluted Ising models, with nearest neighbor antiferromagnetic (AF) bonds, on fcc lattices, \cite{henley,DS} and in (iii) strongly site-diluted Ising models with dipole-dipole interactions, such as in LiHo$_x$Y$_{1-x}$F$_4$. \cite{chicago} We refer to the latter systems as disordered Ising dipolar (DID) systems.\cite{yes0,yes} At least for DID systems, some numerical evidence that is unfavorable for the existence of a phase transition also exists.\cite{yu}  
The divergence of $\xi$ implies the divergence of the so called spin-glass susceptibility $\chi_{sg}$ at $T_{sg}$, where $\chi_{sg}=N^{-1}\sum_{ij}\langle \langle s_is_j \rangle_T ^2 \rangle_q$ and $N$ is the number of spins.

Convincing experimental evidence for the existence of an equilibrium phase transition into the SG phase is  harder to obtain. This is mainly (i) because  very long relaxation processes make equilibrium observations difficult, and (ii) because neither $\xi$ nor $\chi_{sg}$ can be directly observed. Instead, the  SG transition is usually characterized by the nonlinear magnetic susceptibility $\chi_{3}$. \cite{PW} It is defined by
\begin{equation}
m=\chi_1 H+\chi_3H^3+\cdots ,
\label{uno}
\end{equation}
assuming $m(-H)=-m(H)$. Canella and Mydosh\cite{mydosh} were first able to measure  (in gold-iron alloys) huge values of $\chi_3$: $T_{sg}^2\chi_{3}/\chi_1 \sim 10^{5}$ (from here on, we let Boltzmann's constant and  Bohr's magneton equal $1$)  near $T=T_{sg}$.
Later, ${\chi}_{3}$ was  shown \cite{monod} to diverge  in  other canonical SGs as a power of $T-T_{sg}$.
For the EA model,\cite{ea} originally inspired by the discovery of Canella and Mydosh, Chalupa \cite{chalupa} showed long ago that
\begin{equation}
-\chi_3 = T^{-3} (\chi_{sg}-2/3)
\label{cuatro}
\end{equation}
if no field is applied. Thus, the critical behavior of $\chi_3$ and $\chi_{sg}$, which one observes in simulations, can, at least for the EA model, be clearly related to the critical behavior of $\chi_3$, which one observes experimentally. 

The three models we study are governed by the Hamiltonian,
\begin{equation}
{\cal H}=-\frac{1}{2}\sum_{ij}J_{ij}x_is_ix_js_j,
\label{ham}
\end{equation}
where the sum is over all $i$ and $j$ lattice sites, $J_{ij}$ is model specific, $x_i=\pm 1$ is a quenched random variable, and $s_i$ is a ($\pm 1$) Ising spin at site $i$. All sites are occupied with the same probability, $x=\langle x_i\rangle_q$, where the $q$ subscript stands for a quenched average over all site occupancy arrangements.

The aim of this paper is to find how $-\chi_3 $, an experimentally measured quantity, and $\chi_{sg}$, a quantity which is often calculated, are related in site-diluted SGs.
More specifically, numerical results from TMC are sought for (i) Ising spins, with Ruderman-Kittel-Kasuya-Yoshida (RKKY) interactions,\cite{RKKY} which are randomly located on a small fraction of all lattice sites,  (ii) a geometrically-frustrated Ising spin system, mainly, randomly located Ising spins, with nearest neighbor AF interactions, on a $0.4$ fraction of all sites of an fcc lattice, and (iii) DID systems  
on a small fraction of all lattice sites.  The outcome of these calculations is unknown, because
Eq. (\ref{cuatro}) has not been derived for site diluted SGs. On the other hand, $\chi_3$ and $\chi_{sg}$ can exhibit the same critical behavior in site diluted SGs if they and the  EA model belong to the same universality class.
This has been  predicted\cite{bray} to be so for the first of the above three models, but not  so, as far as we know,  for the other two models.\cite{grest,other}

An outline of the paper follows. Details about the procedure we follow in our calculations are given in Sec. \ref{pro}.
For various sizes of each of these systems, we obtain $\xi /L$, $\chi_{sg}$, and $\chi_{3}$.
Data for $\xi /L$ is used to establish the phase transition temperature $T_{sg}$ between the paramagnetic and SG phases. We then compare how $\chi_{sg}$ and $\chi_3$ vary with system size and with temperature in the vicinity of $T_{sg}$. The results obtained for each system are given in each of the subsections of Sec. \ref{results}. Very briefly, these results follow. 
We find that $\chi_3$ approximately follows Eq. (\ref{uno}) in strongly diluted systems of Ising spins with RKKY interactions. 
On the other hand, $-T^3\chi_3$ is a over a couple of orders of magnitude smaller than $\chi_{sg}$ in a ($x=0.4$) site-diluted AF Ising model on an fcc lattice. Nevertheless, both $\chi_3$ and $\chi_{sg}$ appear to have the same critical behavior. Finally, 
in DID systems, $-T^3\chi_3$ and $\chi_{sg}$ seem to diverge similarly at $T_{sg}$. However, $-T^3\chi_3/\chi_{sg}$ varies sharply with  systems' shape.  Taking into account demagnetization effects, we estimate in Sec. \ref{shape} how $-T^3\chi_3$ varies with system shape for  high aspect ratios.
Near the transition temperature, $-T^3\chi_3/\chi_{sg}$  increases from $-T^3\chi_3/\chi_{sg}\sim 10^{-2}$ for cubic shape systems to $-T^3\chi_3/\chi_{sg}\sim 10^{2}$ for very thin long prisms. 
Our conclusions are summarized in Sec. \ref{con}. 

As a byproduct, we have obtained  values for $\eta$ and $T_{sg}$ in these three systems. They are listed in Table II.

From here on, in addition to $k_B=1$, $\mu_B=1$, we let $m=N^{-1}\sum_i\langle\langle s_i \rangle_T\rangle_q$, and assume
spins in all models  point up or down along the $z$-axis, sometimes referred to as the magnetization axis.

\section {Procedure}
\label{pro}
To calculate $\chi_3$, we make use of
\begin{equation}
6\chi_3=   N^{-1}T^{-3}(\langle M^4 \rangle_T
 -3  \langle M^2 \rangle_T ^2),
\label{mn}
\end{equation}
where $M=Nm$, which holds for $H=0$. This equation  follows  from Eq. (\ref{uno}) by (i) repeated differentiation with respect to $H$ of the canonical ensemble average expression for $ \langle m \rangle_T $, and by (ii) letting
$\Delta =0$, where $\Delta =-4 \langle M^3 \rangle  \langle M \rangle +12 \langle M^2 \rangle  \langle M \rangle^2 -6  \langle M \rangle^4$. 
This is justified for finite systems with up-down symmetry if averages are taken over infinite times, since $\langle M^n \rangle =0$ then for all odd $n$. The order in which system sizes and averaging times are taken to infinity is irrelevant for the paramagnetic phase. Equations (\ref{cuatro}) and (\ref{mn}), as well as all the results below are only claimed to hold for $T\geq T_{sg}$. 
 Note that Eq. (\ref{mn}) is valid for each realization of quenched disorder. Chalupa derived Eq. (\ref{cuatro})  from  Eq. (\ref{mn}) for the EA model
by first averaging over all system samples, and noting (i) that both $ \langle M^4 \rangle$ and $\langle M^2 \rangle ^2$ involve sums over four-spin terms, such as $\sum_{ijkl}\langle\langle s_is_js_ks_l\rangle_T\rangle_q$ and $\sum_{ijkl}\langle\langle s_is_j   \rangle_T \langle s_ks_l\rangle_T\rangle_q$, respectively, and (ii) that
any term in which  one or more subindices is unpaired vanishes. To see this, assume the $k$ index is unpaired in either of the two sums over $ijkl$ indices. Now, consider all exchange bonds $J_{km}$ between the $k$-th and any other site. Let's assume the probabilities for $J_{km}$ and $-J_{km}$ for all $m$ while all other exchange constants remain unchanged are equal. (This requires exchange bonds to be  \emph{independently} random.)
It follows that the probabilities for $s_k$ and $-s_k$, for any given configuration of all the other spins, are equal. This is the gist of the proof.  For further details, see Ref. [\onlinecite{chalupa}]. The proof fails for site-diluted systems, because exchange bonds are not then  \emph{independently} random.

We simulate a set of identical systems at temperatures $T_{min}$, $T_{min}+\Delta T$, $T_{min}+2 \Delta T, \ldots T_{max}$ following standard TMC rules. \cite{TMC} We choose $T_{max}\simeq 2.5 T_{sg}$, $T_{min}\sim 0.5T_{sg}$, and all $\Delta T$ such that  at least $30\%$ of all attempted
exchanges between systems at $T$ and $T+\Delta T$ are successful for all $T$.  We let each system equilibrate for a time $\tau_s$ and take averages over an equally long subsequent time $\tau_s$. Time $\tau_s$ satisfies two requirements, (i) that $\langle \langle M\rangle_T^2\rangle_q \ll 0.1 \langle \langle M^2\rangle_T\rangle_q$ obtains for all $T\in[T_{min},T_{max}$], and (ii)
that, systems which start from either random configurations or from (assumed) equilibrium configurations come to the same condition, as specified in Ref. [\onlinecite{solo}], after time $\tau_s$. 
Values of $\tau_s$, of the number of samples $N_s$ with different quenched randomness over which averages were taken, and of the site occupancy rate $x$, are given in Table I.

Periodic boundary conditions are used throughout. For DID systems, we make use of Ewald sums. \cite{ew}

For the correlation length $\xi$, we make use, as has become standard practice,\cite{balle,jorgo,yes,solo} of the original definition,\cite{old} 
\begin{equation} 
\xi^2=\frac {1 } {4 \sin^2  ( k /2)}  { \left[ \frac{ \chi_{sg}   }{ \mid \chi_{sg} (\textbf{k})\mid     }  -1 \right] },
\label{nosa}
\end{equation}
where $\chi_{sg}(\textbf{k})=N^{-1}\sum_{ij} \langle \langle s_i s_j\rangle_T^2 \rangle_q\exp ({i\textbf{k}\cdot\textbf{r}_{ij}})$, and we let ${\bf k} =(2\pi/L,0,0)$. 
Note that $\chi_{sg} (\textbf{k}=0)=\chi_{sg}$ and  that, as $L\rightarrow \infty$,  $\xi /L$ vanishes in the paramagnetic phase, remains finite at $T=T_{sg}$, and either grows without bounds below $T_{sg}$, as in a conventional phase transition, or remains finite as in the XY model in two dimensions. The point where $\xi /L$ curves for various system sizes meet as $T$ decreases defines $T_{sg}$ for us.

All systems we report on below have a common feature: fractional errors for $\chi_3$ are an order of magnitude larger than the ones for $\chi_{sg}$ and for $\chi_1$. Unless otherwise stated, error bars are given for $\chi_3$ and related quantities, but not for $\chi_{sg}$ or $\chi_{1}$, which are all smaller than icons for their data points.

\begin{table}\footnotesize
\caption{The number of samples $N_s$ and the number of Monte Carlo sweeps $\tau_s$ that were taken both for equilibration and for the subsequent averaging times are given in thousands of Monte Carlo sweeps. There are $L^3$ lattice sites in RKKY and FCC systems, but $L\times L\times 2L$ lattice sites for DID systems.  }
\begin{ruledtabular}
\begin{tabular}{|c| r c r  |r c c c|c r c|}
 & & RKKY&  & &&FCC&  & & DID     \\
\hline
L & 6 & 8 & 12 & 4& 6 &8 &12 & 4 &6 &8\\
$\tau_s$ & $10$ & $ 40$ & $400$ &$10$ &$10$ &$40$ &$40$& $10$ &$100$& $3000$\\
$N_s$ & $100$    & $40$ & $10$ & $10$ &$40$ &$10$ &5.6&$100$ & 15 & 4 \\
$x$ & $0.1$  & $0.1$ & $0.1$ & $0.4$ & $0.4$ & $0.4$ & $0.4$ &$0.35$&$0.35$& $0.35$\\
\end{tabular}
\end{ruledtabular}
\end{table}

\section{results}
\label{results}

All results in this section follow from TMC simulations. 

\begin{figure}[!t]
\includegraphics*[width=80mm]{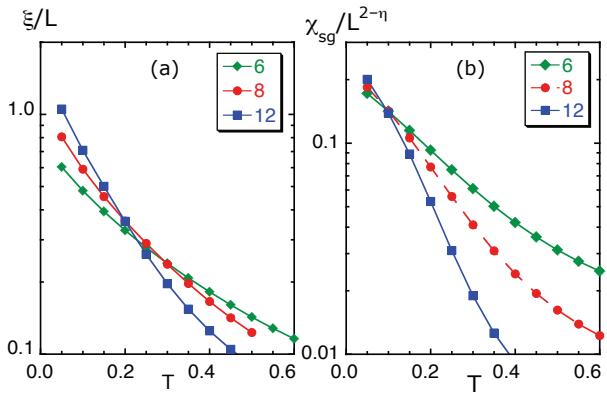}
\caption{(Color online) (a) Semilog plots of $\xi /L$  vs $T$ for ($m=\pm 1$) Ising spins with RKKY interactions, randomly located, on a $0.1$ fraction of all $L^3$ sites. The  numbers in the box are $L$ values. (b) Same as in (a) but for $\chi_{sg} /L^{2-\eta}$ vs $T$, for $\eta=-0.5$. }
\label{primera}
\end{figure}

\subsection
{Spatially disordered Ising spins with RKKY interactions }
The Hamiltonian is given by Eq. (\ref{ham}), with $J_{ij}=\varepsilon_c (\cos kr_{ij} )(a/r_{ij})^3$, 
ie, an RKKY \cite{RKKY} interaction, as in a canonical spin glass. 
We let $ka=2\pi$, where $a$ is a nearest neighbor distance, and $\varepsilon_c$ is an energy in terms of which all temperatures are given in this subsection. 
We let each site is be occupied with $x=0.1$ probability.

Plots of $\xi /L$ vs $T$  for various system sizes are shown in Fig. \ref{primera}a. Not all pairs of curves  cross at the same point. Let $T_{i,j}$ be the temperature where curves for lengths $L_i$ and $L_j$ cross, where $L_1=6$, $L_2=8$ and $L_3=12$. 
Plots of $T_{ij}$ (where $T_{1,2}=0.29$,  $T_{1,3}=0.23$, and $T_{2,3}=0.20$) vs $1/L_iL_j$ fall onto a straight line which extrapolates to $T=0.10$ as $1/L_iL_j\rightarrow 0$. We thus estimate $T_{sg}=0.10$,
and therefore expect $T/x=1.0$ for  $x\lesssim 1$ values, since the $1/r^3$ dependence of the interaction implies\cite{yes} $\xi(x,T)=\xi(T/x)$.

In Fig. \ref{primera}b we can see that   $\chi_{sg}$ seems to grow without bounds with system size at  $T= T_{sg}$. Indeed, we note that $\chi_{sg}\sim L^{2-\eta}$, where $\eta \simeq -0.5$ at $T_{sg}$. This $\eta$ value at the boundary of the range $-0.5\lesssim \eta \lesssim -0.2$ of quoted\cite{katz} values, from Monte Carlo simulations, for the EA model. We note in passing that this model and the EA model have been predicted\cite{bray} to be in the same universality class.

 \begin{figure}[!t]
\includegraphics*[width=80mm]{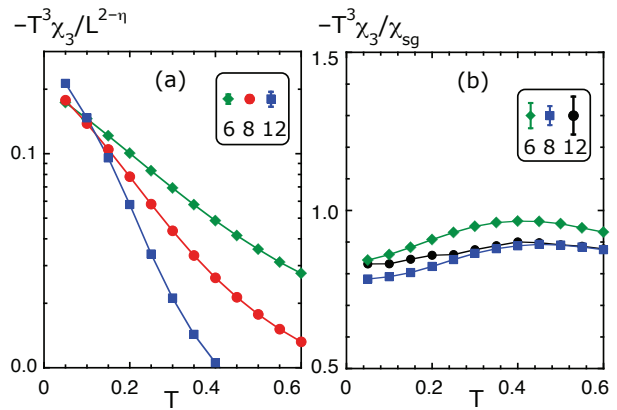}
\caption{(Color online) (a) Semilog plots of $-T^3\chi_3 /L^{2-\eta}$ vs $T$, for $\eta =-0.5$, for ($m=\pm 1$) Ising spins with RKKY interactions, randomly located, on a $0.1$ fraction of $L^3$ sites, for the  values of $L$, which are shown in the box. (b) Same as in (a) but for $-T^3\chi_3/\chi_{sg}$ vs $T$.}
\label{primeru}
\end{figure}

How $-T^3\chi_3$ behaves near $T_{sg}$ is shown in Fig. \ref{primeru}a. It varies with $T$ and with $L$ much as $\chi_{sg}$ does in Fig. \ref{primera}b. The plots  shown in  Fig. \ref{primeru}b are consistent with $-T^3\chi_3 \sim \chi_{sg}$. (The value $\eta = -0.5$ follows from the plots shown in Fig. \ref{primera}b, not from any fitting of $\chi_3$ to any desired behavior.)

More significantly, $-T^3\chi_3/ \chi_{sg}$ appears to approach  a \emph{smooth} function of temperature in the neighborhood of $T=T_{sg}$ as $L\rightarrow \infty$. This is the basis for the main  conclusion of this section, namely, that  $-T^3\chi_3$  and $ \chi_{sg}$ have the same critical behavior.

\subsection
{Site-diluted AF Ising model on a fcc lattice}
Each site of a fcc lattice is occupied with a ($\pm 1$) Ising spin with a $0.4$ probability. The Hamiltonian is given by Eq. (\ref{ham}), with  $J_{ij}=-J$ if $i$ and $j$ are  nearest neighbors but $J_{ij}=0$ otherwise.    A $0.4$ occupancy rate is roughly midway between the lowest value $x=0.195$ for percolation \cite{perc} in fcc lattices and the transition point, $x\simeq 0.75$, between SG and AF phases.  \cite{henley} All temperatures are given in terms of $J$.

\begin{figure}[!t]
\includegraphics*[width=80mm]{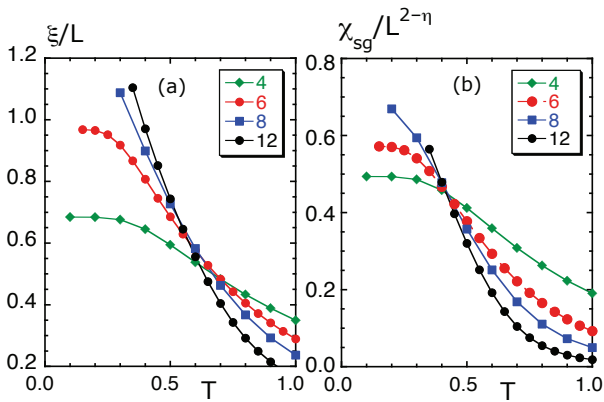}
\caption{(Color online) (a) Plots of $\xi /L$  vs $T$ for a ($x=0.4$) site-diluted AF Ising model on an fcc lattice of $L\times L\times L$ sites. The  numbers in the box are $L$ values. (b) Same as in (a) but for $\chi_{sg} /L^{2-\eta}$ vs $T$, for $\eta=-0.5$.}
\label{tercera}
\end{figure}

Monte Carlo results for this model are shown in Figs. \ref{tercera}a and \ref{tercera}b. We note in Fig. \ref{tercera}a that the crossing point between pairs of $\xi /L$ curves drifts leftward as their $L$-values increase. 
As for Fig.  \ref{primera}a, let $T_{i,j}$ be the temperature where curves for lengths $L_i$ and $L_j$ cross, where $L_1=4$, $L_2=6$, $L_3=8$ and $L_4=12$. 
A second degree polynomial fit to a plot of  $T_{ij}$ (where $T_{1,2}=0.70$,  $T_{1,3}=0.67$,  $T_{1,4}=0.62$, $T_{2,3}=0.62$, $T_{2,4}=0.57$, and $T_{3,4}=0.53$) vs $1/L_iL_j$ gives $T_{ij}\rightarrow 0.4$ as $1/L_iL_j\rightarrow 0$. We thus estimate $T_{sg}=0.4(1)$,  in agreement, within errors, with the values found for $x=0.4$ in Ref. [\onlinecite{henley}]. Plots of $\chi_{sg}/L^{2-\eta}$ vs $T$ are shown in Fig.  \ref{tercera}b for $\eta=-0.5$. This is the best value of $\eta$ to have  $\chi_{sg}/L^{2-\eta}$ curves for various values of $L$ cross at $T_{sg}$.
This value of $\eta$ is, within errors, in agreement with the value found for $x=0.4$ in Ref. [\onlinecite{henley}]. 

\begin{figure}[!t]
\includegraphics*[width=80mm]{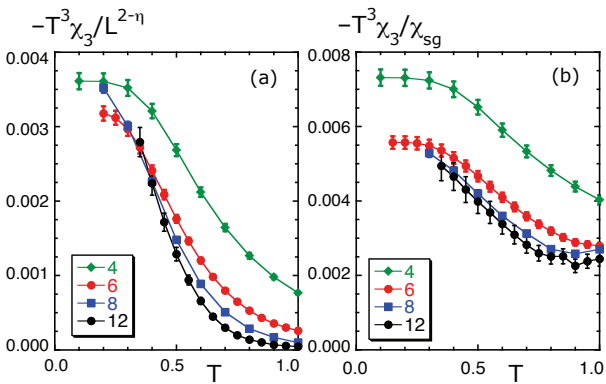}
\caption{(Color online) (a) Plots of $-T^3\chi_3/L^{2-\eta}$ vs $T$, for $\eta =-0.5$, for a ($x=0.4$) site-diluted AF Ising model on an fcc lattice of $L\times L\times L$ sites. The numbers in the box are $L$ values. (b) Same as in (a) but for $-T^3\chi_3 /\chi_{sg} $  vs $T$.}
\label{fcc2}
\end{figure}

Plots of $-T^3\chi_3/L^{2-\eta}$ vs $T$, with $\eta =-0.5$, are shown in Fig.  \ref{fcc2}a. The  $\eta = -0.5$ value is taken from the plots of $\chi_{sg}/L^{2-\eta}$ vs $T$, not from any fitting of $\chi_3$ to any desired behavior. The curves in  Figs.  \ref{tercera}b and  \ref{fcc2}a are somewhat different, but all  curves for $L=6,8$ and $12$ in both figures do cross, \emph{within errors}, at the same temperature, $T_{sg}=0.4$. 

In Fig.  \ref{fcc2}b   we notice that $-T\chi_3\ll \chi_{sg}$, which differs markedly from what might have been expected from the behavior of the EA model (and from the above results for  SGs with RKKY interactions).  More significantly, we observe $-T\chi_3/ \chi_{sg}$ 
is, within errors, independent of $L$ for the largest values of $L$, and appears to go into a smooth function of $T$, near $T_{sg}$, as $L\rightarrow \infty$. This suggests that, in the thermodynamic limit, both quantities have the same critical behavior. 

\begin{table}\footnotesize
\caption{Values for $T_{sg}$ follow from crossing (or merging) points of $\xi /L$ curves for various values of system linear size, $L$. 
Values of $\eta$ are assigned so that $\chi_{sg}/L^{2-\eta}$ curves for various $L$ values cross at $T_{sg}$. Errors in $\eta$ follow from errors in $T_{sg}$. 
As explained in Ref. [\onlinecite{yes}], $T_{sg}$  can be obtained for DID systems (for all $x\lesssim 0.5$ in sc lattices\cite{yes} and $x\lesssim 0.25$ in\cite{bel} LiHo$_x$Y$_{1-x}$F$_4$)  from the $T_{sg}$ value given below, making use of
$T_{sg}\propto x$. Similarly for RKKY interactions and all $x\lesssim 0.1$}
\begin{tabular}{|c|c |c|c|}
\hline
&  RKKY  &  FCC  &   DIDs  \\
\hline
$T_{sg}$  & $0.10(4)$  for  $x= 0.1$  & 0.4(1) for $x=0.4$ & $0.35(4)$ for $x= 0.35$   \\
$\eta $  & $-0.5(4)$ & $ -0.5(2)$  & $0.0(3)$ \\
\hline
\end{tabular}
\end{table}

\subsection{Spatially disordered ($\pm 1$) Ising dipoles}

Here we consider disordered Ising dipolar (DID) systems in sc lattices. We let each site be occupied, with a $0.35$ probability, by a ($\pm 1$) spin. All spins point up and down, along the $z$-axis.
The Hamiltonian is given by Eq. (\ref{ham}), with
\begin{equation}
J_{ij}=h_d
\frac{a}{r_{ij}^3}^3\left( 3
\frac{z_{ij}^2}{r_{ij}^2} -1\right),
\label{T}
\end{equation} 
where $ r_{ij}$ is the distance between $i$ and $j$ sites, $z_{ij}$ is the $z$ component of $ r_{ij}$, $h_d$ is an energy, and $a$ is the
SC lattice constant. 

Let's first recall that, despite some earlier numerical evidence to the contrary, \cite{yu} more recent calculations point to the existence of a phase transition between the paramagnetic and SG phases in  diluted Ising dipolar systems \cite{yes0,yes} at $T_{sg}/x\simeq 1$ for all $x\lesssim 0.5$.  In addition,\cite{solo} $\chi_{sg}\sim L^{2-\eta}$ at $T=T_{sg}$, where $\eta\simeq 0$. 

We deal with the \emph{magnetic} susceptibility here which, as is well known,  depends on the shape of the system.\cite{morrish}  For this reason we study numerically $L\times L\times nL$ shaped prism systems for various values of $n$, that is, square-base prisms with a $1:n$ aspect ratios.

\begin{figure}[!t]
\includegraphics*[width=80mm]{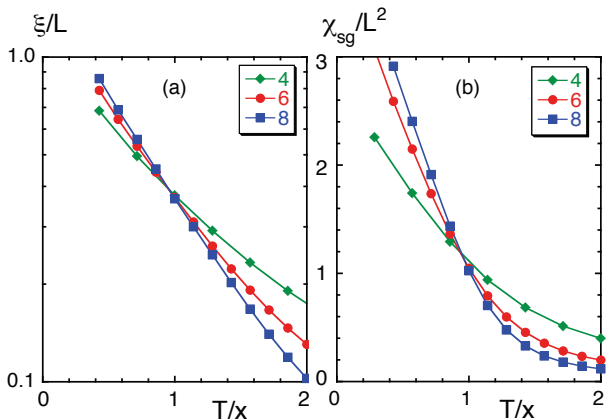}
\caption{(Color online) (a) Semilog plots of $\xi /L$  vs $T/x$ for DID systems on a $0.35$  fraction of  all $L\times L\times L_z$ sites, where $L_z=2L$. The numbers in the box are  $L$ values. (b) Same as in (a) but for $\chi_{sg} /L^{2}$ vs $T/x$. Error bars are smaller than icons for all data points.}
\label{did1}
\end{figure}

Plots of $\xi /L$  vs $T/x$ are shown in Fig. \ref{did1}a  for $n=2$. Curves for three different values of $L$ are observed to cross at $T/x\simeq 1.0$. This transition temperature value is in agreement with  the result found in Ref. [\onlinecite{yes}] for $n=1$, mainly, that $T_{sg}/x\simeq 1.0$ for all $x\lesssim 0.5$.

Plots of $\chi_{sg}/L^{2}$ vs $T/x$ are shown  in Fig. \ref{did1}b. All curves are observed to cross at $T/x=0.95$. This is approximately as in Ref. [\onlinecite{yes}].

\begin{figure}[!t]
\includegraphics*[width=80mm]{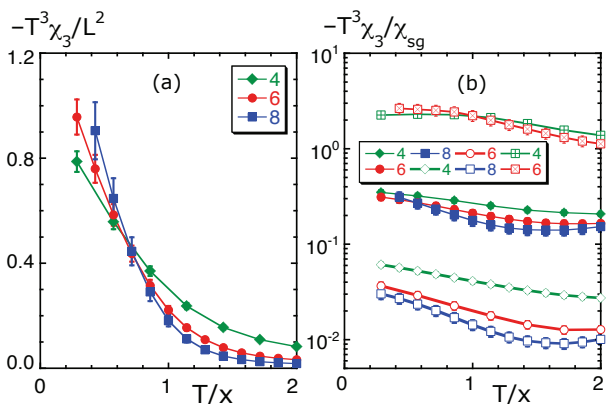}
\caption{(Color online) (a) Plots of $-T^3\chi_3 /L^2$  vs $T/x$ for DID systems on a $0.35$  fraction of  all $L\times L\times L_z$ sites, where $L_z=2L$.  (b) Plots of $-T^3\chi_3 /\chi_{sg}$ vs $T/x$ for three $1:n$ aspect ratios, $n=4$ ($\boxplus$ and $\boxtimes$)  for the two top curves, $n=2$ (full icons) for the  three curves in the middle, and $n=1$ (empty icons) for the three lower curves.  Error bars hardly protrude from icons. For both (a) and (b), the numbers in the box are the values of $L$.}
\label{did2}
\end{figure}

We now turn our attention to the  \emph{magnetic} susceptibility. Plots of $-T^3\chi_3/L^2$ vs $T/x$ are shown in Fig. \ref{did2}a for systems of various sizes. These plots resemble the ones for $\chi_{sg}$ in Fig. \ref{did1}b, but note that the crossing points  are not quite at the same temperature in the two figures.


In order to better compare $-T^3\chi_3$ and $\chi_{sg}$, we plot in Fig. \ref{did2}b the ratio $-T^3\chi_3/\chi_{sg}$ vs $T/x$ for systems of various sizes  with $1:4$, $1:2$ and $1:1$ aspect ratios. For a $1:4$ aspect ratio, only data points for systems with $4\times 4\times 16$ and $6\times 6\times 24$ sites appear in Fig.  \ref{did2}b. A larger system with the same aspect ratio would have taken a prohibitively long computer time to run. Let $L_\bigstar$ be a system length such that 
$-T^3\chi_3/\chi_{sg}$ is approximately size independent if $L \gtrsim L_\bigstar$. Clearly, $L_\bigstar \simeq 4$ and $6$ for $n=4$ and $2$, respectively, in Fig. \ref{did2}b.  For $n=1$,  $L_\bigstar \simeq 8$ seems likely.  This would be in accordance with the expectation that $-T^3\chi_3$ and $\chi_{sg}$ have the same critical behavior in DID systems, independently of aspect ratio. 

Questions  about the sharp variation of  $-T^3\chi_3/\chi_{sg}$ with respect to aspect ratio naturally arise. What is the  asymptotic behavior of  $-T^3\chi_3/\chi_{sg}$?
This is hard to foresee from the data plots shown in Fig. \ref{did2}b. To proceed much further numerically  is impractical. The next section is devoted to this question.

\section{Variation of $\chi_3$ with aspect ratio in DID systems}
\label{shape}

In this section we derive an approximate equation for the variation of $\chi_3$ with shape in DID systems.

Consider two systems of the same shape and size. In  system $f$, all spin pairs interact. In the other system, system $t$, dipole-dipole interactions are truncated.  In $t$, each spin interacts only with spins that lie within a long thin cylinder centered on it, whose axis is parallel to the system's $z$-axis. The radius of the cylinder  need not be more than a couple of nearest neighbor distances, but  its length must be much longer than its radius. Let's furthermore assume that both systems are homogeneous, that is, all sites are occupied ($x=1$). Now, we know from Ref. [\onlinecite{penultimo}] that if both systems are in thermal equilibrium, and external magnetic fields $H_t$ and $H_f$ are applied to systems $t$ and $f$, respectively, such that $m$ is the same in both systems, then
\begin{equation}
H_f=H_t-{\lambda}_nm,
\label{h}
\end{equation}
where $n$ comes from $f$ system's $1:n$ aspect ratio. Equation (\ref{h}) holds because the only effect of the \emph{un}truncated portion of all dipole-dipole interactions in $f$ is to give the so called \emph{demagnetizing} field, ${-\lambda}_nm$. For \emph{dipolar} prisms of $1:n$ aspect ratio,  a scaling expression, such as\cite{privman} $m=t^{-\beta}f(Ht^{-\beta\delta})$ must therefore be replaced by 
\begin{equation}
m=t^{-\beta}f[(H-\lambda_nm)t^{-\beta\delta}].
\label{m}
\end{equation}
where $H_f$ has been replaced by $H$.

Taking the derivative of Eq. (\ref{m}) with respect to $H$ [or, more simply, of Eq. (\ref{h}) with respect to $m$] gives
\begin{equation}
\frac{1}{\chi_1(n)}  =  \frac {1} {\chi_1(\infty)}+\lambda_n,
\label{chi1}
\end{equation}
where $\chi_1(n)$ is the linear susceptibility of a prism with a $1:n$ aspect ratio, and, clearly, $dm/dH_t =\chi_1(\infty)$.
This is the 
well known equation,\cite{morrish}
that experimentalists\cite{chicago,quilliam} often use in order to do away with demagnetization effects, and thus relate $\chi_1(n)$, the measured susceptibility, to ${\chi_1(\infty)}$.

Taking the  $d/dH_t$ derivative of Eq. (\ref{chi1}) gives
\begin{equation}
\left( 1+\lambda_n \frac{dm}{dH_t}\right )     \frac {1}{\chi_1^2(n )} \frac{d\chi_1(n )}{dH_n} =  \frac {1}{\chi_1^2(\infty )} \frac{d\chi_1(\infty )}{dH_t} ,
\end{equation}
where we have used 
$dH_n/dH_t = 1+\lambda_n dm/dH_t$, which follows from Eq. (\ref{h}). 
We next (i) take the $d/dH_t$ derivative of the above equation, (ii) let $H_t=H_n=0$, and (iii)
let $d\chi_1(\infty )/dH_t=0=d\chi_1(n)/dH_n$, by up-down symmetry. The result is easily cast into
\begin{equation}
\chi_3(n )=\frac{\chi_3(\infty)}{[1+\lambda_n\chi_1(\infty)]^4}, 
\label{chi3}
\end{equation}
which is the desired expression relating $\chi_3(n)$ and $\chi_3(\infty )$.

\begin{figure}[!t]
\includegraphics*[width=80mm]{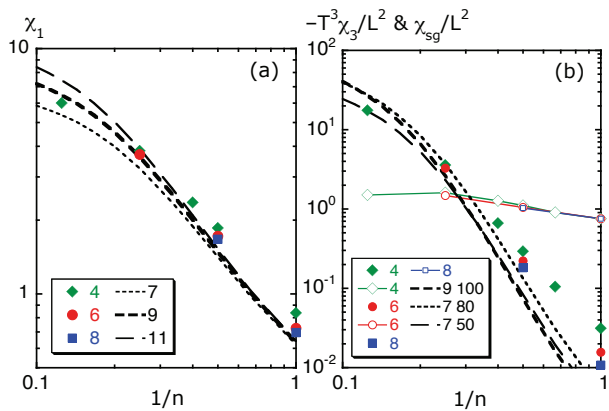}
\caption{(Color online)  (a) Plots of $\chi_1$  vs $1/n$ for DID systems at $T=T_{sg}$, on a $0.35$  fraction of  all $L\times L\times nL$ sites.  The shown numbers are  $L$  values for data points from Monte Carlo calculations. The dashed lines follow from Eq. (\ref{chi1}), assuming the three  values for $\chi_1 (\infty )$ that are shown in the box. (b) Same as in (a) but  for (full icons) $-T^3\chi_3/L^2$ and (open icons) $\chi_{sg}/L^2$. The dashed lines follow from Eqs. (\ref{chi1}) and  (\ref{chi3}), and  the shown pairs of values, such as $9$ and $100$, are for $\chi_1(\infty )$ and $\chi_3(\infty )$, respectively.}
\label{theory}
\end{figure} 
\begin{table}\footnotesize
\caption{$\lambda_n$ values, in terms of $h_d$, for some $n$ in the $[0.5, 7]$ range.  For $n\geq 8$, $\lambda_n\simeq 8/n^2$. For an $x$ site occupancy rate, $\lambda_n\rightarrow x\lambda_n$. }
\begin{ruledtabular}
\begin{tabular}{|c| l l l  l l l l l l l l|}
$n$ & $0.5$ & $1$ & $1.5$ & $2$ &$2.5$ &$3$ &$4$ &$5$& $6$ &$7$& \\
$\lambda_n$ & 7.419& $4.189$    & $2.503$ & $1.611$ & $1.107$ &$0.802$ &$0.471$ &0.308&$0.217$ & 0.160 & \\
\end{tabular}
\end{ruledtabular}
\end{table}

Equations  (\ref{chi1}) and (\ref{chi3}) enable us to calculate how $\chi_3(n)$ varies with $n$ if we know $\chi_1$ and $\chi_3$ for at least an aspect ratio each, as well as $\lambda_n$. A  list of (easily computed) $\lambda_n$ values for several values of $n\in [0.5,7]$, as well as a functional relation for all $n\geq 8$, are given in Table III.
Since the only effect of the long-range portion of all dipole-dipole interactions in $f$ is to give the  demagnetizing field,\cite{penultimo} ${-\lambda}_nm$,
we can calculate all $\lambda_n$ assuming a fully occupied lattice in which all spins point up. Since only a single state comes into the calculation, no Monte Carlo simulation is necessary.
This enables us to calculate $\lambda_n$ for very large  systems.
The fact that only an $x$ fraction of lattice sites are occupied in  site diluted SGs is \emph{approximately} taken into account by letting $\lambda_n \rightarrow x\lambda_n$ everywhere. Inhomogeneities in SGs are thus neglected.

Equation (\ref{chi1}) gives the three dashed lines shown in Fig. \ref{theory}a for $\chi_1(\infty)=7, 9$ and $11$. With two of these values,   we obtain from Eq. (\ref{chi3})
the three curves  for $\chi_3(n)$ shown in Fig. \ref{theory}b for values of $\chi_3(\infty )$. These curves do not fit the data points too well. On the other hand,  a good fit  for small system sizes should not be expected. We can nevertheless conclude with some confidence that $\chi_3(n)$ does not diverge as $n\rightarrow \infty$. Indeed, $\chi_3(\infty )$ is most likely within the $(50,120)$ range. Furthermore, observation of Fig. \ref{theory}b indicates that $\chi_3(n)/\chi_{sg}$ at $T=T_{sg}$ varies over three or four orders of magnitude as system shape varies from cubic to infinitely thin needle-like. 

It is perhaps worth pointing out that 
\begin{equation}
\chi_3(n)=   \frac{ \chi_3(\infty) }{ \chi_1^4(\infty ) } \chi_1^4(n) 
\label{xyz}
\end{equation}
follows immediately from Eqs. (\ref{chi1}) and (\ref{chi3}) after Eq. (\ref{chi1}) is cast into $\chi_1(n)=\chi_1(\infty )/[1+\lambda_n\chi_1(\infty )]$. Equation (\ref{xyz}) implies that $\chi_3$ sweeps over  four times as many decades as $\chi_1$ does (compare Figs. \ref{theory}a and \ref{theory}b) as $n$ varies.

Finally, note that the classical or quantum nature of DID systems does not play any role in this section. It does not matter either whether a transverse field is applied, because  it does not affected up-down symmetry. These equations can therefore be applied, as an illustration, to Li$_1-x$Ho$_x$Y$_4$, under a transverse field, as in Ref. [\onlinecite{bel}], where  $T^2\chi_3/\chi_1\sim 1$ was observed on a $1.6 \times 16 \times 5$ mm$^3$ sample. Values of $\chi_1$ and $\chi_3$ that would be some $3$ and $100$ times larger, respectively, for a long thin needle-like sample can be read off from Figs. \ref{theory}a and \ref{theory}b.

\section{conclusions}
\label{con}

By the tempered Monte Carlo method\cite{TMC} we have tested whether the relation $-T^3\chi_3=\chi_{sg}-2/3$, which is known\cite{chalupa} to hold for the Edwards-Anderson model, also holds for several site-diluted spin glasses of ($\pm 1$) Ising spins, with (i) RKKY interactions, (ii) antiferromagnetic interactions between nearest neighbor spins on fcc lattices, and (iii) dipole-dipole interactions. As a byproduct, we have obtained the values of $\eta$ and $T_{sg}$ that  are listed in Table II.

We have found $-T^3\chi_3\sim \chi_{sg}$  to hold, for Ising spins with RKKY interactions 
occupying a $0.1$ fraction of all lattice sites. More significantly, $-T^3\chi_3/ \chi_{sg}$ appears to be (i) independent of linear system size,  within errors, and (ii) a smooth function of temperature near $T_{sg}$. This suggests $-T^3\chi_3 $ and $ \chi_{sg}$ have the same critical behavior.
Since the RKKY interaction decays as the inverse of the cube of the distance, these results must hold for lower values of $x$ if the temperature is scaled with $x$.

We have found $-T^3\chi_3$ to be over two orders of magnitude smaller than $\chi_{sg}$  for Ising spins, with antiferromagnetic interactions, on a ($x=0.4$) site diluted fcc lattice. 
Our results are, however, consistent with identical critical behavior of these two quantities. 

In  DID systems the TMC data (see Fig. \ref{did2}b) are consistent with $\chi_3$ and $\chi_{sg}$ diverging in the same manner as $T \rightarrow T_{sg}$ from above. 
The sharp variation of $-T^3\chi_3/\chi_{sg}$ with aspect ratio, which can be observed in Fig. \ref{did2}b, is noteworthy. How this comes about from demagnetization effects is explained in Sec. \ref{shape}. In it, relations are derived which together with  data points coming from TMC simulations (see Fig. \ref{theory}b) give rough estimates of $-T^3\chi_3$ at or near $T=T_{sg}$. 
We find  $-T^3\chi_3/\chi_{sg}$ varies, as shown in Fig. \ref{theory}b from $-T^3\chi_3/\chi_{sg}\sim 10^{-2}$ for cubic shapes to $-T^3\chi_3/\chi_{sg} \sim 10^{2}$ for long thin needles.

Our results for DID systems with an $x=0.35$ site occupancy rate can be generalized to smaller values of $x$. As discussed in Ref. [\onlinecite{yes}],
any physical quantity $f$ satisfies $f(x,T)=f(T/x)$ for $x$ quite smaller than $x_c$, the critical concentration above which there is magnetic order at low temperature (e.g., $x_c\simeq  0.65$ for sc lattices\cite{yes} and\cite{bel} $x_c\simeq 0.25$ for LiHo$_x$Y$_{1-x}$F$_4$).

\acknowledgments
I am grateful to Juan J. Alonso for helpful remarks after reading the manuscript. 
Funding Grant FIS2009-08451, from the Ministerio de Ciencia e Innovaci\'on of Spain, is acknowledged

\end{document}